\documentclass[useAMS,usenatbib]{mn2e}
\usepackage{graphicx}
\usepackage{amssymb}
\newcommand{\xmmn}{{\it XMM-Newton~\/}}
\newcommand{\asca}{{\it ASCA~\/}}

\newcommand{\suzaku}{{\it Suzaku~\/}}

\def\H0{{\rm ~km~s^{-1}~Mpc^{-1}}}

\def\la{\mathrel{\hbox{\rlap{\hbox{\lower4pt\hbox{$\sim$}}}{\raise2pt\hbox{$<$}}}}}
\def\ga{\mathrel{\hbox{\rlap{\hbox{\lower4pt\hbox{$\sim$}}}{\raise2pt\hbox{$>$}}}}}
\def\d25{D$_{25}$}

\def\deg{\hbox{$^\circ$}}

\title[Exploring the disc/jet interaction in 4C~+74.26] {Exploring the
  disc/jet interaction in the radio-loud quasar 4C~+74.26 with Suzaku}

\author[J.~Larsson, et al.]  {\parbox[]{6.0in}
  {J. Larsson~$^1$\thanks{E-mail: jlarsson@ast.cam.ac.uk},
    A. C. Fabian~$^1$, D. R. Ballantyne~$^2$ and G. Miniutti~$^{1,3}$\\
    \footnotesize {\it $^1$Institute of Astronomy, University of
      Cambridge, Madingley Road, Cambridge CB3 0HA\\
      $^2$Department of Physics, The University of Arizona, 1118
      E. 4th Street, Tucson, Arizona 85721 USA\\
      $^3$Laboratoire Astroparticule et Cosmologie (APC), UMR 7164, 10
      Rue A.  Domon et L. Duquet, 75205 Paris, France}}}

\date{Accepted 2008 May 21.  Received 2008 May 15; in original form
  2008 April 8}

\pagerange{\pageref{firstpage}--\pageref{lastpage}}
\pubyear{2008}

\begin{document}

\maketitle

\label{firstpage}

\begin{abstract}
  We report on a 90~ks \suzaku observation of the radio-loud quasar
  4C~+74.26. The source was observed in its highest flux state to
  date, and we find that it brightened by about 20 per cent during the
  observation. We see evidence of spectral hardening as the count rate
  increases and also find that the rms variability increases with
  energy up to about 4~keV. We clearly detect a broadened Fe line but
  conclude that it does not require any emission from inside about
  50~$r_{\rm g}$, although a much smaller inner radius cannot be ruled
  out. The large inner radius of our best fit implies that the inner
  disc is either missing or not strongly illuminated. We suggest that
  the latter scenario may occur if the power-law source is located
  high above the disc, or if the emission is beamed away from the
  disc.

 \end{abstract}
            
\begin{keywords}
  galaxies: active -- galaxies: individual: 4C~+74.26 -- X-rays:
  galaxies.
\end{keywords}

\section{Introduction}

Only a minority of active galactic nuclei (AGN) harbour relativistic
radio-jets. The mechanism that drives the jet production in these
radio-loud objects is unknown and constitutes an important issue in
the study of accreting black holes. As both jets and X-ray emission
originate close to the central black hole, it is clear that insight
into this problem can be gained by comparing the X-ray properties of
radio-loud and radio-quiet objects (see Ballantyne 2007 for a recent
review).

Such studies of radio-loud and radio-quiet AGN, as well as Galactic
black holes in their different states, have shown that radio-loud
objects generally have weaker reflection features (as measured by the
Fe~K$\alpha$ line and Compton hump) and narrower Fe lines than their
radio-quiet counterparts (see e.g Wo\'{z}niak et al. 1998; Eracleous,
Sambruna \& Mushotzky 2000; Grandi, Malaguti \& Fiocchi 2006 for AGN
and Fender, Belloni \& Gallo 2004 for Galactic black
holes). Explanations for these findings include highly ionized
accretion discs (Ballantyne, Ross \& Fabian 2002) and dilution by jet
emission (e.g. Page et al. 2005), but the usually favoured
interpretation is that the thin accretion disc in radio-loud objects
is truncated at large radii and replaced by an advection-dominated
accretion flow (ADAF). The latter explanation is especially appealing
in terms of jet production, as ADAFs are thought to naturally lead to
bipolar outflows (Narayan \& Yi 1994, 1995). While ADAFs are expected
to exist within objects with very low accretion rates
($\dot{M}/\dot{M}_{\rm Edd} \lesssim 0.01$, Rees et al. 1982), such as
low-luminosity AGN and Galactic black holes in their low/hard states,
they are unlikely to be present in radio-loud quasars, which generally
have higher accretion rates.

Over the past few years there have been several reports of radio-loud
objects, most of them quasars, where the presence of very broad Fe
lines suggest that the accretion disc extends close or all the way in
to the innermost stable orbit. Examples include the radio-loud AGN
4C~+74.26 (Ballantyne \& Fabian 2005), 3C~273 (T\"{u}rler et
al. 2006), PG~1425+267 (Miniutti \& Fabian 2006), 3C~109 (Allen et
al. 1997; Miniutti et al. 2006) and 3C~120 (Kataoka et al. 2007), as
well as the Galactic black hole GX 339-4 in its low/hard state (Miller
et al. 2006). These findings strongly indicate that a truncated thin
disc is not a necessary condition for jet formation, and in particular
not appropriate for objects with relatively high accretion rates. In
this paper we will present the analysis of a new \suzaku observation
of one of these objects, 4C~+74.26.

4C~+74.26 is a broad line radio galaxy (BLRG) located at $z=0.104$
(Riley et al. 1989). Its radio luminosity is on the boarder of FRII
and FRI but its morphology is clearly that of an FRII object (Riley et
al. 1989). A one-sided jet which is at least 4~kpc long has been
observed with the VLA (Riley \& Warner 1990), and on parsec scales
with VLBI (Pearson et al. 1992). Based on the flux-limit for a counter
jet, Pearson et al. (1992) find that the inclination of the source
axis must be less than $49^{\circ}$ to the line of sight. The
bolometric luminosity of the source is around $2\times 10^{46}\
\rm{erg\ s^{-1}}$, making it a low-luminosity quasar (Woo \& Urry
2002). The same authors also estimate a black hole mass of $4\times
10^9\ M_{\tiny{\sun}}$, which means that the source is operating at an
Eddington ratio of $\sim 0.04$.

4C~+74.26 was first detected in the X-rays in the ROSAT All-Sky Survey,
but the first observation that allowed a detailed spectral analysis
was carried out with \asca in 1996 (Brinkmann et al. 1998). Brinkmann
et al. (1998) detected an Fe line with $EW\approx 100$~eV but could
not determine if the line was broadened. The continuum could be well
fitted with a model consisting of a power law and cold reflection
(with a large reflection fraction of $\sim$ 6), modified by absorption
at low energies. The \asca data were later re-analysed by Sambruna,
Eracleous \& Mushotzky (1999), who favoured a double power-law model
for the continuum, in which the hard power law was attributed to
emission from a jet. For the double power-law model the Fe line was
found to be broad ($\sigma \sim 0.6$~keV) and the equivalent width
higher ($\sim 200$~eV).

The spectrum of a subsequent {\emph BepposSAX} observation could be
described with a similar model as the one presented by Brinkmann et
al. (1998), but with a significantly smaller reflection fraction of
around 1 (Hasenkopf, Sambruna \& Eracleous 2002). An \xmmn observation
later revealed a broad Fe line, which implied an inner disc radius
close to the innermost stable circular orbit for a maximally spinning
black hole (Ballantyne \& Fabian 2005). At low energies the \xmmn
spectrum showed evidence for both cold and warm absorption (Ballantyne
2005). 4C~+74.26 has also been detected in the \emph {Swift} BAT and
\emph{INTEGRAL} surveys (Tueller et al. 2007 and Bodaghee et al. 2007,
respectively).

In this paper we present the results of a 158~ks ($\sim 90$~ks net
exposure) \suzaku observation of 4C~+74.26 performed in October 2007.
We also carry out joint fits with the \xmmn observation performed in
February 2004, and briefly discuss the data from the \emph {Swift} BAT
and \emph{INTEGRAL} catalogs. Our main goal is to try and determine
the inner radius of the accretion disc. This paper is organised as
follows: Section \ref{observations} describes the observations and the
data reduction, Section \ref{variability} describes the variability
and Section \ref{spectra} describes the spectral analysis. We finally
discuss our results in Section \ref{discussion} and present our
conclusions in Section \ref{conclusions}.

\section{Observations and data reduction}
\label{observations}

4C~+74.26 was observed by \suzaku between 2007 October 28--30 for a total
duration of 158 ks. Event files from version 2.1.6.16 of the {\it
  Suzaku} pipeline processing were used and spectra were extracted
using {\scriptsize XSELECT}.

For each XIS, source spectra were extracted from circular regions of
4.3 arcmin radius centred on the source (which was observed off-axis
in the HXD nominal position). Background spectra were extracted from
two circular regions with the same total area as the source region,
avoiding the chip corners with the calibration sources. Response
matrices and ancillary response files were generated for each XIS
using {\scriptsize XISRMFGEN } version 2007-05-14 and {\scriptsize
  XISSIMARFGEN} version 2007-09-22. The ARF generator should account
for the hydrocarbon contamination on the optical blocking filter
(Ishisaki et al. 2007).

The net exposure time of all three XIS detectors is 92~ks and the
2--10~keV count rates are $0.753\pm 0.003\ \rm{counts\ s}^{-1}$(XIS0),
$0.806\pm 0.003\ \rm{counts\ s}^{-1}$ (XIS1) and $0.854\pm 0.003\
\rm{counts\ s}^{-1}$ (XIS3). Because of the very similar spectra of
the two front-illuminated (FI) detectors (XIS0 and XIS3), their
spectra were co-added for the spectral analysis, and their response
files were combined accordingly.

For the HXD/PIN detector, a model for the non-X-ray background (NXB)
was provided by the HXD team. Source and background spectra were
constructed from identical good time intervals, and the exposure time
of the background spectrum was increased by a factor of 10 (to account
for the fact that the background model was generated with 10 times the
actual count rate in order to minimize the photon noise). The source
spectrum was corrected for dead time, leaving a total net exposure
time of 86~ks. The response file appropriate for the HXD aim point was
used (ae\_hxd\_pinhxnome4\_20070914.rsp)

The total PIN count rate over the 14--30 keV energy range is $0.3860
\pm 0.0021\ \rm{counts\ s}^{-1}$, compared to $0.2743 \pm 0.0005\
\rm{counts\ s}^{-1}$ for the background. Since the background is so
much brighter than the source, the accuracy of the background model is
very important for our results. According to Mizuno et al. (2007), the
one-sigma uncertainly of version 2 background models is around 3 per
cent. It is often possible to determine the accuracy of the background
model in a given observation more precisely by comparing it with the
night earth spectrum. However, in the case of 4C~+74.26 the PIN field
of view was never completely obscured by the earth, and such a
comparison is not possible. Some information on the accuracy of the
background can also be obtained by comparing the light curves of the
source and the background. This will be carried out in section
\ref{variability}.

The background model discussed above does not include the contribution
from the cosmic X-ray background (CXB). In order to account for this
we simulate the spectrum of the CXB and add it to the spectrum of the
non-X-ray background. For the simulation of the CXB spectrum we use a
model of the form $8 \times 10^{-4} (E/1\ \rm{keV})^{-1.29}\
exp{(-E/40\ \rm{keV})}$, which is based on the HEAO-A1 spectrum,
renormalized to the HXD field of view. The CXB count rate in the
14--30~keV band is about 6 per cent of the count rate of the total
background (NXB + CXB).

\subsection{The \xmmn observation}
\xmmn observed 4C~+74.26 on 2004 February 06 for a total duration of
34~ks. The data were reduced as described in Ballantyne (2005), using
the \xmmn Science Analysis System version 7.1.0. We use only the data
from the EPIC pn camera, which, after removing intervals of background
flaring, has a total good exposure time of 29~ks.

\section{Variability}
\label{variability}

\subsection{light curves}
\label{lightcurves}
\begin{figure}
\begin{center}
\rotatebox{270}{\resizebox{!}{80mm}{\includegraphics{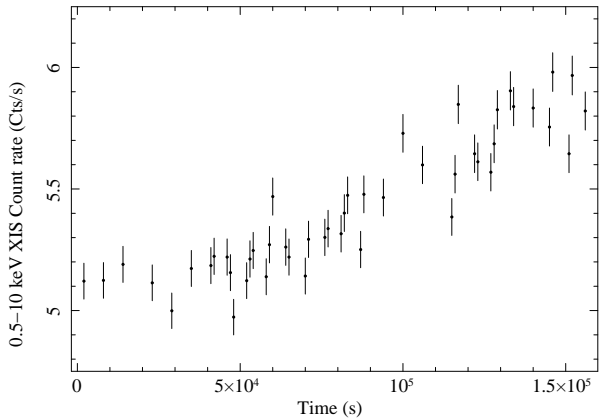}}}
\caption{\small{0.5-10 keV XIS (0+1+3) light curve on a 1 ks
    time-scale, using only fully-exposed bins.}}\label{xislc}
\end{center}
\end{figure}
\begin{figure}
\begin{center}
  \rotatebox{270}{\resizebox{!}{80mm}{\includegraphics{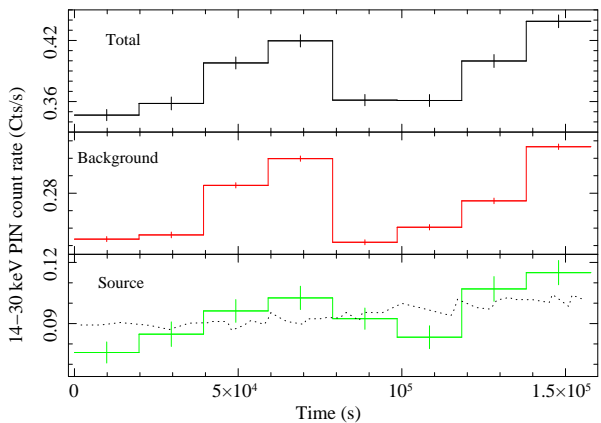}}}
  \caption{\small{14--30~keV PIN light curves on a 20~ks
      time-scale. Panels from top to bottom show the total light
      curve, the light curve of the NXB model, and the source light
      curve (where both the NXB model and the constant contribution
      from the CXB have been subtracted).  As a reference the bottom
      panel also shows the XIS light curve (dotted line), rescaled to
      the mean PIN count rate. The background-subtracted light curve is
      clearly correlated with the background model, suggesting that
      the background has been underestimated.}}\label{pinlc}
\end{center}
\end{figure}

Fig. \ref{xislc} shows the 0.5--10~keV XIS light curve from the
\suzaku observation of 4C~+74.26. The light curve was obtained by
adding the light curves of all three XIS detectors, using
fully-exposed 1 ks bins. The source brightens by about 20 per cent
during the observation. We also note that the 2--10~keV flux has
gradually increased by about a factor of two since the source was
observed by \asca in 1996, as summarized in Table \ref{longterm}. The
flux in the soft band in the \asca observation was also found to be
two times higher than in a ROSAT observation in 1993 (when the source
was in the field of view of an observation targeted at the cataclysmic
variable VW Cep, Brinkmann et al. 1998). The 2--10~keV flux in the
\suzaku observation is $\rm{{3.24\times 10^{-11}\ erg\ cm^{-2}\
    s^{-1}}}$, corresponding to a luminosity of $\rm{{8.80\times
    10^{44}\ erg\ s^{-1}}}$.

In Fig \ref{pinlc} we show 14--30~keV PIN light curves on a 20~ks
time-scale. Specifically, we compare the total light curve, the light
curve of the NXB model, and the light curve of the source.  The source
light curve was obtained by subtracting both the NXB model and the
constant contribution from the CXB. As a reference we also show the
XIS light curve, rescaled to the mean PIN count rate. The light
curves of the background model and the source are clearly correlated
(the cross-correlation function peaks at a value of 0.82 at 0 time
lag), indicating that the background model has been underestimated.

\begin{table}
\begin{center}
\begin{tabular}{llrc}
  \hline\hline \\ [-4pt]
Obs. date & Mission &\multicolumn{1}{c}{2-10~keV flux} &
  \multicolumn{1}{c}{Reference} \\[2pt]
& & \multicolumn{1}{c}{(${\rm 10^{-11}\ erg\ cm^{-2}\ s^{-1}}$)}& \\ [4pt]
  \hline\\[-6pt]
 2007-10-28 & \suzaku & 3.24 & (1)\\
2004-02-06 & \xmmn & 2.43 & (2)\\
1999-05-17 & \emph{BepppoSAX} & 1.41    & (3)\\
1996-09-09 & \asca & 1.69 & (4)\\
  \hline
\end{tabular}
\caption{\label{longterm}\small{2--10~keV fluxes of
    4C~+74.26. References: (1) This paper, (2) Ballantyne \& Fabian
    (2005), (3) Hasenkopf et al. (2002), (4) Sambruna et al. (1999).}}
\end{center}
\end{table}

\subsection{Hardness ratios}

\begin{figure}
\begin{center}
\rotatebox{270}{\resizebox{!}{80mm}{\includegraphics{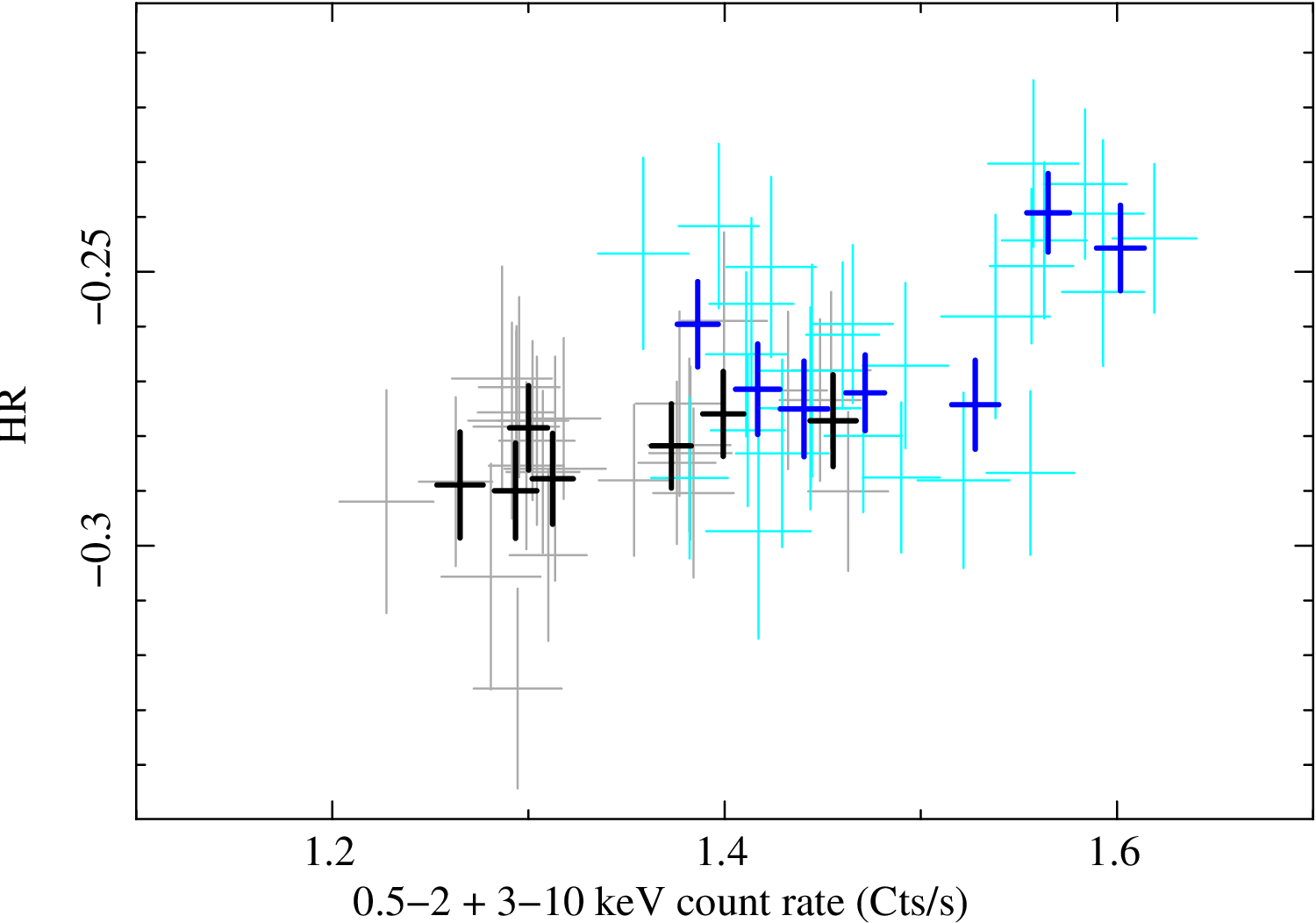}}}
{\vspace{0.25cm}}
\rotatebox{270}{\resizebox{!}{80mm}{\includegraphics{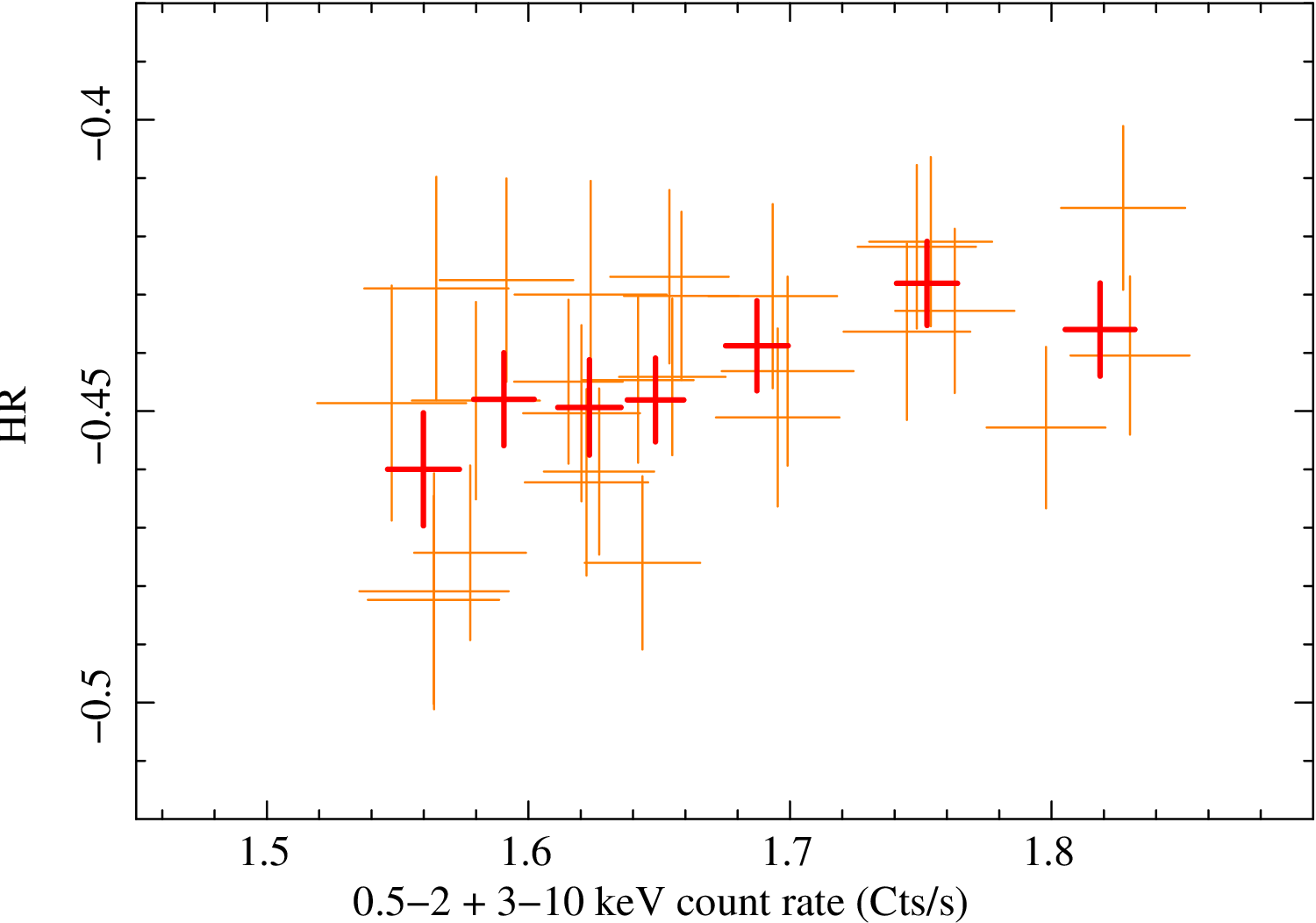}}}
\caption{\small{hardness ratios as a function of count rate for XIS 0
    and 3 (black and blue crosses, upper panel) and XIS1 (lower
    panel). The hardness ratio is defined as HR=(H-S)/(H+S), where S
    is the count rate in the $0.5-2$ keV band and H is the count rate
    in the $3-10$ keV band. The hardness ratios are calculated based
    on $\sim6$~ks spectra. The thin crosses represent un-binned data
    points and the thick ones show the mean of data points binned
    along the x-axis. Each bin contains four original data points,
    apart from the last bin which contains three. For all three XIS we
    see that the spectrum tends to harden as the count rate
    increases.}}
\label{hratios}
\end{center}
\end{figure}

In order to characterise the spectral variability of the source we
start by calculating hardness ratios, $\rm{HR=H-S/H+S}$, where we take
H to be the count rate in the 3--10~keV band and S to be the count
rate in the 0.5--2~keV band. The hardness ratios for all three XIS,
calculated in orbital length bins of 5760~s, are plotted as a function
of the $\rm{H+S}$ count rate in Fig. \ref{hratios}. In order to reduce
the scatter in the plot we binned consecutive points along the x-axis
so that each bin comprises four original data points (apart from the
last bin which contains three data points). Un-binned and binned data
are shown as thin and thick crosses, respectively.

For all three XIS we see that the spectrum tends to harden as the
count rate increases, although within a fairly large spread. In
agreement with this, we find that a positive correlation is a better
fit than a constant for all three data sets. According to an F-test,
the improvement in the fit for un-binned (binned) data is significant
at 91 (93) per cent (XIS0), 99 (99) per cent (XIS1) and 98 (87) per
cent (XIS3). The fact that the positive correlation is highly
significant only in XIS1 is presumably due to the higher effective
area at low energies in this detector, which reduces the error on the
count rate in the soft band.

\subsection{The rms spectrum}
\begin{figure}
\begin{center}
\rotatebox{270}{\resizebox{!}{80mm}{\includegraphics{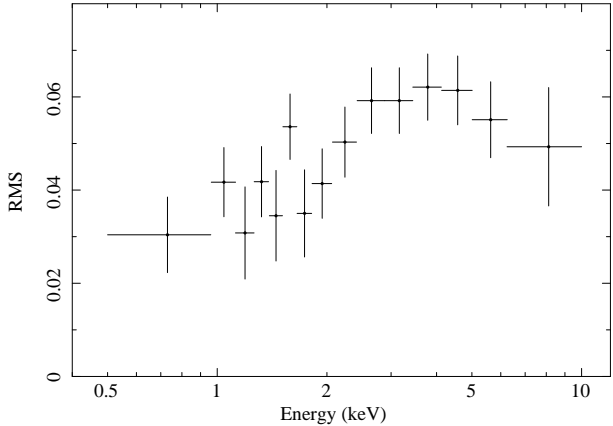}}}
\caption{\small{Rms spectrum of the entire observation calculated on
    the orbital time scale ($\sim6$~ks). The spectrum was calculated
    as an average of the rms spectra of the three XIS
    detectors.}}\label{rms}
\end{center}
\end{figure}

We next consider the root-mean-squared (rms) spectrum, which, for a
given time-scale, shows the fractional variability as a function of
energy. The techniques for calculating rms spectra are described in
e.g. Edelson et al. (2002) and Vaughan et al. (2003).  Fig. \ref{rms}
shows the rms spectrum of 4C~+74.26, calculated as an average of the
rms spectra of the three XIS detectors, on the orbital time-scale of
5760~s. The errors from the Poisson noise were calculated following
Vaughan et al. (2003). The rms variability clearly increases with
energy up to about 4~keV, at which point it flattens out and shows
some evidence of decreasing.

\section{Spectral analysis}
\label{spectra}

In the spectral analysis of the \suzaku data we will concentrate on
the 2--10~keV (XIS) and 14--30~keV (PIN) energy ranges.  We find that
the agreement between the FI XIS is very good in the 2--10~keV range,
and we therefore add their spectra, as previously mentioned in section
\ref{observations} and recommended in the \suzaku Data Reduction
Guide. The spectrum of XIS1, on the other hand, differs significantly
from the FI XIS spectra in the Fe~K$\alpha$ region (the line energy
and width are inconsistent with the FI XIS values), and we do not
include it in our spectral analysis. Because of inconsistencies
between the XIS detectors at low energies (presumably due to the
modelling of the contamination), we do not extend the spectral
analysis below 2~keV at this time. We note, however, that the
low-energy spectrum is characterized by absorption in excess of the
Galactic value (see Fig. \ref{ratio}), in agreement with previous
observations. The absorption is not strong enough to affect the
spectrum above 2~keV.

Below we will first consider the 2--10~keV FI XIS spectrum and then
include the high-energy PIN data. We will also investigate the
high-energy properties of the source using the spectra from
\emph{Swift} and \emph{INTEGRAL}.  As a last step we will perform
joint fits to the \suzaku and \xmmn 2--10~keV spectra. All spectral
fits were performed using {\scriptsize XSPEC} version 11.3.2aj. Errors
on model parameters are quoted at the 90 per cent confidence level and
energies of spectral features are quoted for the rest frame of the
source. All fits include Galactic absorption fixed at $\rm {N_H = 1.19
  \times 10^{21}\ cm^{-2}}$ (Dickey \& Lockman 1990).

\subsection{The 2--10~keV FI XIS spectrum}
\begin{figure}
\begin{center}
\rotatebox{270}{\resizebox{!}{80mm}{\includegraphics{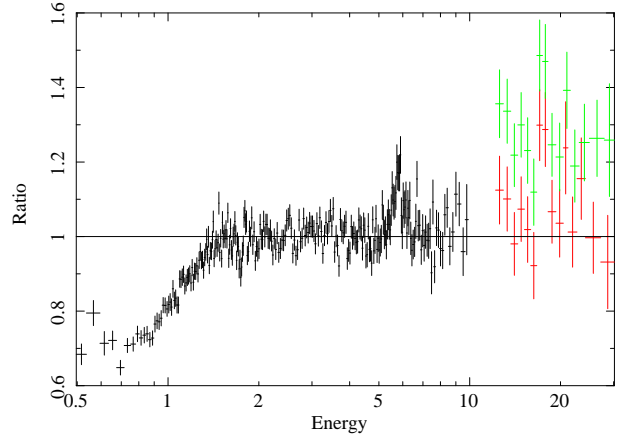}}}
\caption{\small{FI XIS + PIN spectrum as a ratio to a power law
    (modified by Galactic absorption), fitted over the 2--4 and
    7--10~keV energy ranges.  The red and green data points show the
    PIN data assuming a background model of $\pm3$ per cent (the
    one-sigma accuracy of v2 background models).  The study of the PIN
    light curves in section \ref{lightcurves} showed that the
    background has probably been underestimated, suggesting that the
    real value should be closer to the red data points. The low-energy
    data clearly reveal the presence of absorption in excess of the
    Galactic value.}}\label{ratio}
\end{center}
\end{figure}

\begin{figure}
\begin{center}
\rotatebox{270}{\resizebox{!}{80mm}{\includegraphics{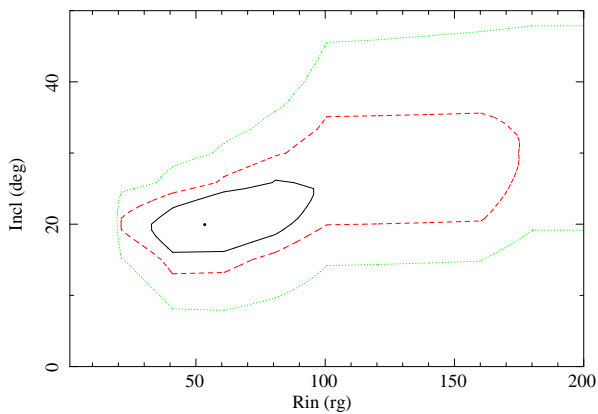}}}
\caption{\small{Confidence contours for $r_{\rm{in}}$ and $i$ in the
    model consisting of a power law and a Laor line. The black
    (solid), red (dashed) and green (dotted) lines represent 1, 2 and
    3 sigma confidence, respectively. The best fit is shown as a black
    dot.}}\label{cont}
\end{center}
\end{figure}

In order to look for the presence of a broad Fe line we start by
fitting the spectrum with a power law, excluding the 4--7~keV energy
band, where the line emission should dominate. Fig. \ref{ratio} shows
the data as a ratio to the power-law fit (which has $\Gamma=1.81\pm
0.01$), clearly revealing that a broadened line is present. The fact
that the line is broadened is confirmed when we add a Gaussian line to
the power-law model. If we let the width of the line be a free
parameter we find an excellent fit of $\chi^2=1510$ for 1509 degrees
of freedom (d.o.f.), compared to $\chi^2/\rm{d.o.f.} = 1546/1510$ if
we force the line to be narrow ($\sigma=1$~eV). The broad line has
$\sigma=0.24^{+0.08}_{-0.07}$~keV and $EW=85^{+20}_{-19}$~eV. The
results of these fits as well as all other spectral fits discussed in
this section are presented in Table \ref{results1}.

Broad Fe lines are often found to have a narrow component originating
from distant matter. To check for such a component in 4C~+74.26 we add
a narrow line at 6.4~keV to the model with the broad Gaussian. This
improves the fit by only $\Delta \chi^2 =-2$ for 2 fewer degrees of
freedom, and the narrow line is found to have a very low equivalent
width of $10^{+15}_{-10}$~eV. We thus conclude that a narrow component
to the Fe line is not required.

We next replace the broad Gaussian with a Laor line (Laor 1991), which
models the Fe line emission from an accretion disc around a Kerr black
hole. We fix the energy of the line at 6.4~keV, the outer radius of
the disc at $r_{\rm out} = 400\ r_{\rm g}$ (where $r_{\rm g}=GM/c^2$ is the
gravitational radius), and the emissivity index of the disc at $q=3$
(the emissivity follows the form $\epsilon \propto r^{-q}$, where $r$
is the radius of emission, and $q=3$ is expected from gravitational
energy release). This leaves the inner radius of emission ($r_{\rm
  in}$), the inclination of the disc ($i$) and the normalization of
the line as free parameters. We find an excellent fit of
$\chi^2/\rm{d.o.f.} = 1514/1509$ with $r_{\rm in}=53^{+119}_{-28}\
r_{\rm g}$ and $i=20^{+8}_{-5}\deg$.

The relatively low inclination obtained in this fit implies that the
source is extremely large (of the order of 3~Mpc). This is still
smaller than the largest radio galaxy observed to date (3C 236 is
$\sim4.5$~Mpc, Saripalli \& Mack 2007) but larger than the majority of
the known radio galaxies.  Assuming that the size of 4C~+74.26 is
around 2~Mpc, which is more typical for a giant radio galaxy, the
inclination should be around 33$\deg$. If we fix the inclination at
this value we obtain a fit which is only slightly worse
($\chi^2/\rm{d.o.f.} = 1518/1510$), but which has a significantly
larger $r_{\rm in}$ of $177^{+223}_{-77}\ r_{\rm g}$. The relationship
between the inclination and $r_{\rm in}$ can be seen in
Fig. \ref{cont}, which shows the confidence contours for these
parameters in the fit where the inclination was left free. We note
that the three-sigma confidence contour remains below $50\deg$, in
agreement with the upper limit on the inclination of $49\deg$ set by
Doppler boosting arguments (Pearson et al. 1992).

\begin{table*}
\begin{center}
\begin{tabular}{lrrrrrrrrrrr}
\hline\hline \\ [-4pt]
Model & \multicolumn{1}{c}{$\Gamma$} & \multicolumn{1}{c}{${\rm E_{G/L}}$}
& \multicolumn{1}{c}{$\sigma_{{\rm G}}$} & \multicolumn{1}{c}{${\rm EW_{G/L}}$} 
&  \multicolumn{1}{c} {${\rm E_{nl}}$} &  \multicolumn{1}{c} {${\rm EW_{nl}}$} 
& \multicolumn{1}{c}{$r_{\rm{in}}$} & \multicolumn{1}{c}{$i$} 
& \multicolumn{1}{c} {Fe} & \multicolumn{1}{c} {$\xi$}  
& \multicolumn{1}{c}{$\chi^2$/dof} \\[4pt]  
\hline\\[-6pt]

pl + nl & $1.80^{+0.01}_{-0.01}$ & & & & $6.40^{+0.05}_{-0.04}$ & $31^{+8}_{-8}$ 
& & & & & 1546/1510\\

pl + G & $1.82^{+0.01}_{-0.02}$ & $6.38^{+0.06}_{-0.05}$ & $0.24^{+0.08}_{-0.07}$ 
&  $85^{+20}_{-19}$ & & & & & & & 1510/1509 \\

pl + G + nl &  $1.82^{+0.01}_{-0.01}$ &  $6.38^{+0.09}_{-0.09}$ 
& $0.30^{+0.12}_{-0.10}$  &  $78^{+28}_{-23}$ & $6.42^{+0.09}_{-0.10}$ 
& $10^{+15}_{-10}$ & & & & & 1508/1507 \\

pl + Laor & $1.81^{+0.01}_{-0.01}$ & $6.40^{\rm{f}}$ & & $70^{+13}_{-16}$ & & 
& $53^{+119}_{-28}$ & $20^{+8}_{-5}$ & & & 1514/1509\\

pl + Laor + nl &  $1.82^{+0.01}_{-0.01}$ &  $6.40^{\rm{f}}$ & & $116^{+51}_{-51}$ 
& $6.40^{+0.05}_{-0.05}$ & $25^{+8}_{-9}$ & $1.24^{\rm{f}}$ & $40^{+4}_{-6}$ & & 
& 1530/1508 \\

pl + blr*ref & $1.84^{+0.03}_{-0.03}$ & & & & & & $68^{+332}_{-25}$ 
& $21^{+18}_{-6}$  & $1.8^{+8.2}_{-0.8}$ & $30^{+34}$ & 1514/1506 \\

pl + blr*ref + nl & $1.87^{+0.05}_{-0.04}$ & & & & $6.40^{+0.05}_{-0.05}$ 
& $27^{+9}_{-10}$ & $1.24^{\rm{f}}$ & $42^{+4}_{-5}$ &  $1.6^{+1.0}_{-0.6}$ 
& $30^{+4}$  & 1528/1507 \\

\hline
\end{tabular}
\caption{\label{results1}\small{Fits to the 2--10~keV \suzaku FI XIS
    spectrum. The model components are pl = power law, nl = narrow
    Gaussian emission line ($\sigma$ fixed at 1~eV), G = Gaussian
    emission line, Laor = Laor model, blr = {\scriptsize KDBLUR}
    (relativistic blurring kernel), ref = {\scriptsize REFLION}
    (reflection model). $\Gamma$ is the photon index of the power law,
    ${\rm E_{G/L}}$ is the energy of the broad Gaussian or Laor line
    (in keV), $\sigma_{{\rm G}}$ is the width of the Gaussian line (in
    keV), ${\rm EW_{G/L}}$ is the equivalent width of the broad
    Gaussian or Laor line (in eV), ${\rm E_{nl}}$ is the energy of the
    narrow line (in keV), ${\rm EW}_{{\rm nl}}$ is the equivalent
    width of the narrow line (in eV), $r_{\rm{in}}$ is the inner
    radius of the disc (in gravitational radii), $i$ is the
    inclination of the disc to the line of sight (in degrees), Fe is
    the iron abundance of the disc (relative to solar), $\xi$ is the
    ionization parameter of the disc (in ${\rm erg\ cm^{-2}\
      s^{-1}}$). Superscript f indicates that the parameter was
    frozen. The following parameters were always fixed: $\sigma$ of
    the narrow line (1~eV), the outer radius of the disc (400
    gravitational radii), the emissivity index of the disc (3). All
    fits also included absorption fixed at the Galactic value.}}
\end{center}
\end{table*}

\begin{figure*}
\begin{center}
\rotatebox{270}{\resizebox{!}{80mm}{\includegraphics{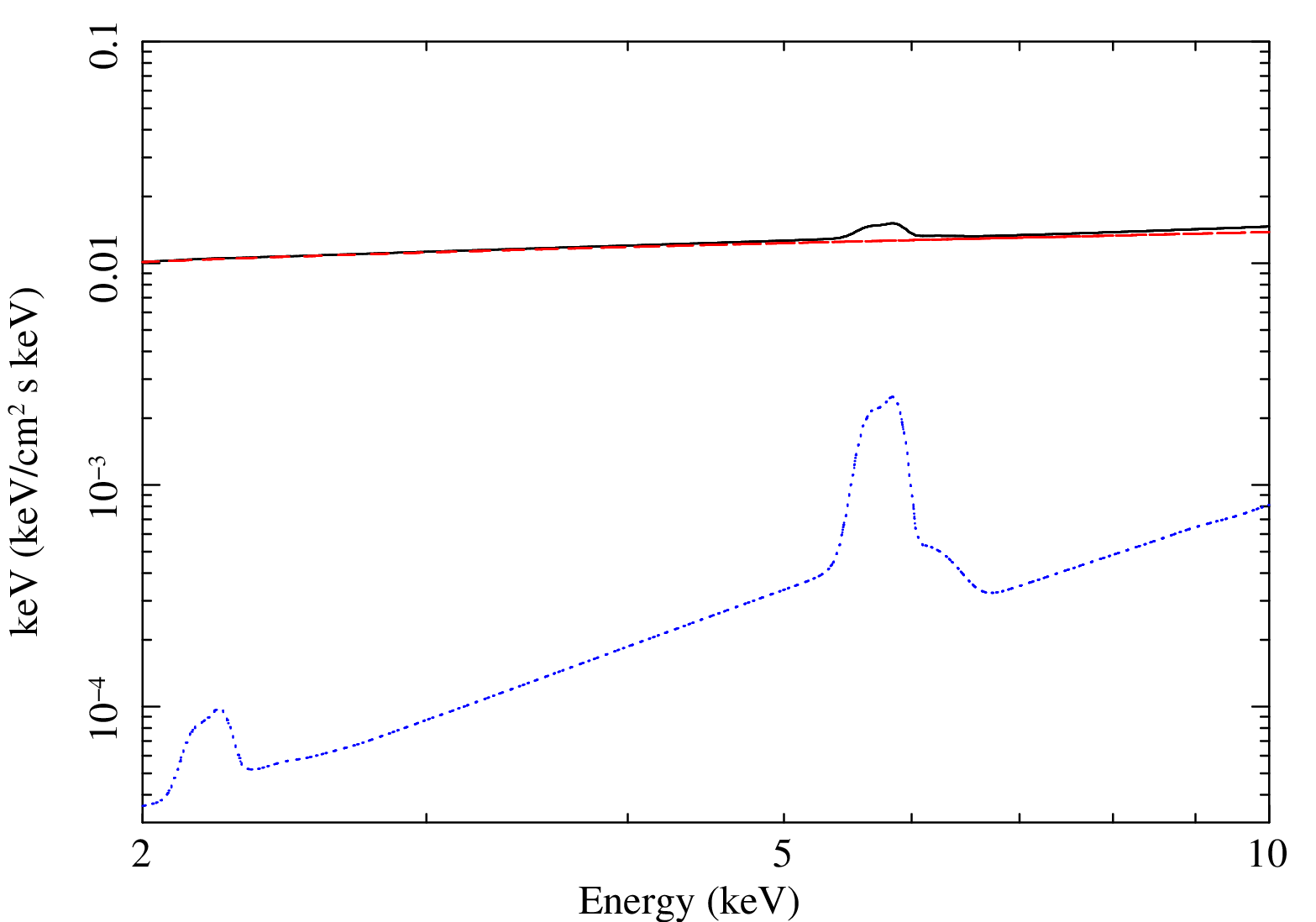}}}
{\hspace{0.15cm}}
\rotatebox{270}{\resizebox{!}{80mm}{\includegraphics{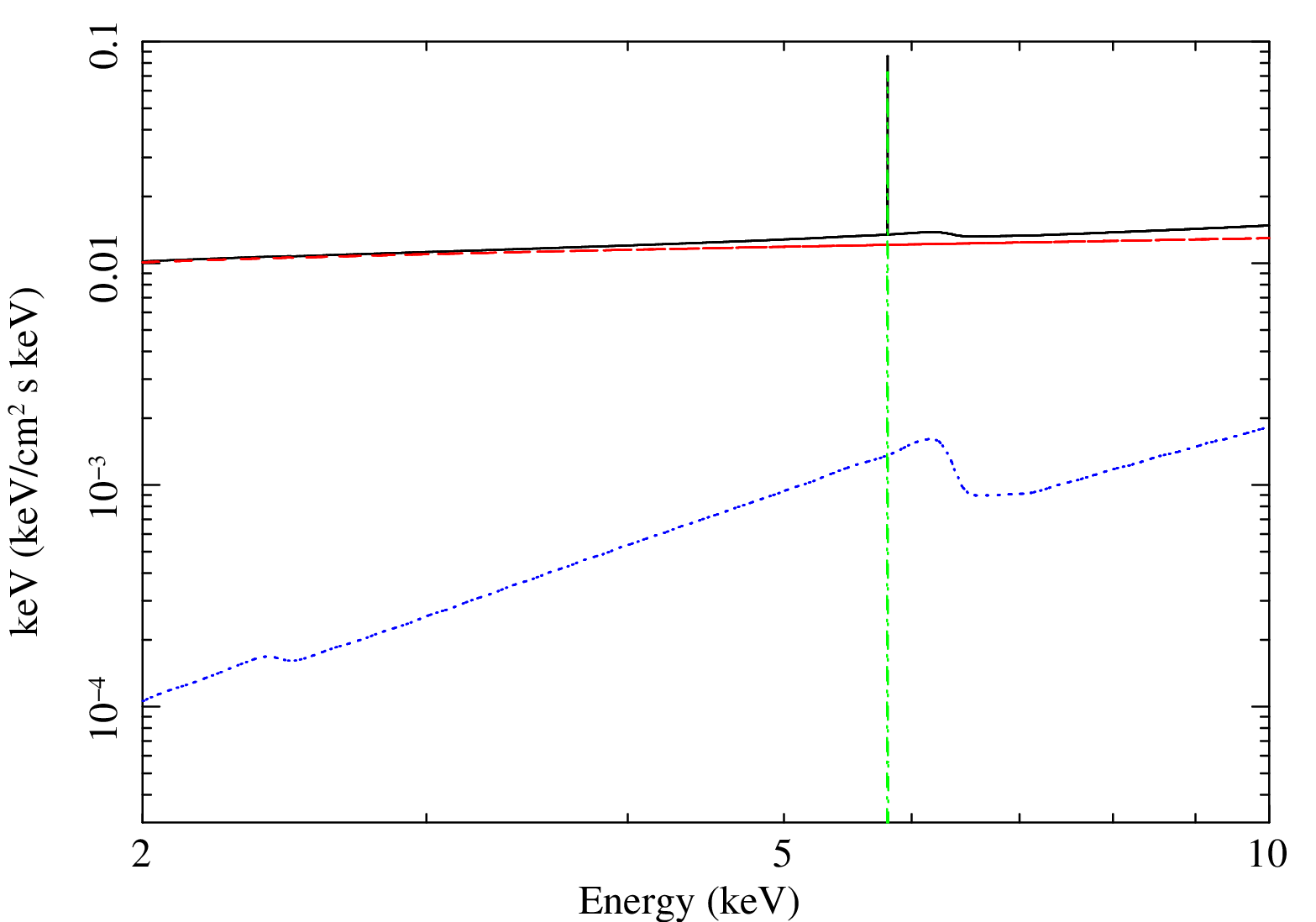}}}
{\vspace{0.15cm}}
\rotatebox{270}{\resizebox{!}{80mm}{\includegraphics{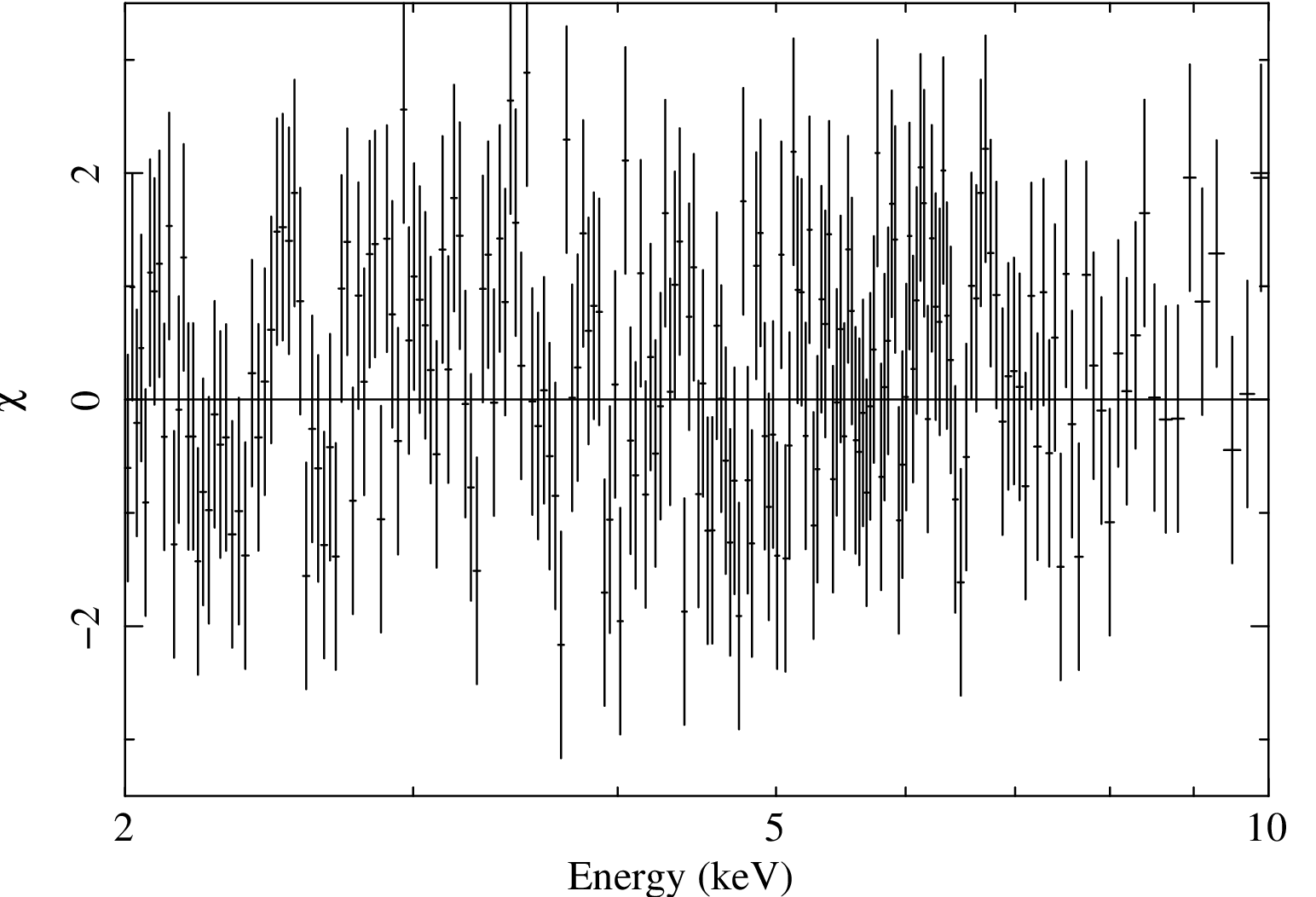}}}
{\hspace{0.15cm}}
\rotatebox{270}{\resizebox{!}{80mm}{\includegraphics{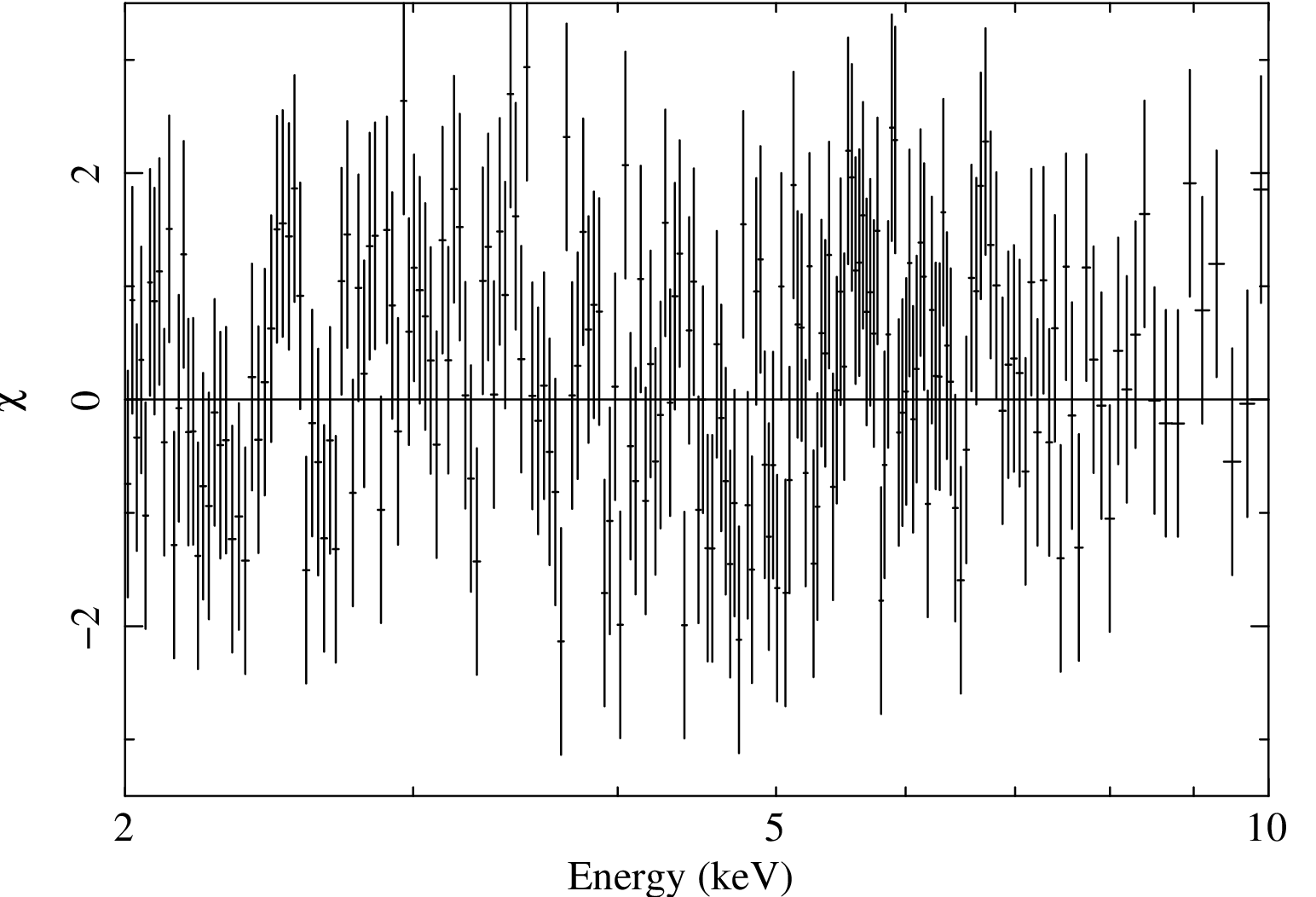}}}
\caption{\small{Reflection fits to the 2--10~keV \suzaku FI XIS
    spectrum. The top panel shows the best-fitting models together
    with their components, and the bottom panel shows the residuals to
    the fit. The model in the left panel consists of a power law and
    reflection from a disc truncated around $70\ r_{\rm g}$. The model shown
    in the right panel consists of a power law, reflection from a disc
    that extends all the way in to $1.24\ r_{\rm g}$, and a narrow
    Fe~K$\alpha$ line.  Black (solid) lines represent the total model,
    red (dashed) lines the power law, blue (dotted) lines the reflection
    model, and the green (dashed-dotted) line the narrow line.}}
\label{refmodels}
\end{center}
\end{figure*}

The relatively large values obtained for $r_{\rm in}$ in the fits
described above suggest that the inner disc in 4C~+74.26 is either not
present or for some reason unable to produce strong fluorescent Fe
emission.  In order to check if a fit where the Fe emission is
produced in the inner regions of the disc can be obtained, we refit
the data with $r_{\rm in}$ fixed at $1.24\ r_{\rm g}$ (the innermost stable
circular orbit of a Kerr black hole). We find that a narrow 6.4~keV
line has to be added to the model in order to obtain an acceptable fit
for this scenario. The quality of the fit is worse than for the case
of a free $r_{\rm in}$ ($\Delta \chi^2 =+15$ for one less d.o.f) but
we note that the combination of the small $r_{\rm in}$ and the narrow
line results in a higher inclination of $40^{+4}_{-6}\deg$.

In order to also model the reflection spectrum associated with the Fe
line, we next replace the Laor line with the {\scriptsize REFLION}
model by Ross \& Fabian (2005), which models the emission from a
constant-density accretion disc. The parameters of the model are the
Fe abundance, the ionization parameter $\xi$, the photon index of the
incident power law and the normalization. In order to account for the
relativistic effects in the vicinity of the black hole, the
reflection model is convolved with the relativistic blurring kernel
{\scriptsize KDBLUR}, which is derived from the code by Laor (1991)
(the model parameters are the inner and outer radius of the disc, the
inclination and the emissivity index). We leave all parameters free to
vary apart from the outer radius of the disc and the emissivity index,
which we fix at $r_{\rm out} = 400\ r_{\rm g}$ and $q=3$, respectively.

We find a best fit of $\chi^2/\rm{d.o.f.} = 1514/1506$ with $r_{\rm
  in}=68^{+332}_{-25}\ r_{\rm g}$ and $i=21^{+18}_{-6}\deg$, which is very
similar to our best-fitting Laor model. As in the case of the Laor
model, the fit only becomes slightly worse if the inclination is fixed
at a larger value. We are also still able to obtain an acceptable fit
with $r_{\rm in}$ fixed at 1.24$\ r_{\rm g}$ if a narrow line at 6.4~keV is
included. As before, this fit is somewhat worse ($\Delta \chi^2 =+13$
for one less d.o.f) and has a higher inclination ($42^{+4}_{-5}\deg$).
In both reflection fits the ionization parameter of the disk is $30\
\rm{erg\ cm\ s^{-1}}$, i.e nearly neutral, and the Fe abundance is
about 2 times solar. Both models are plotted in Fig. \ref{refmodels}
together with the fit residuals. It is clear that the first model
(with the larger $r_{\rm in}$ and lower inclination) is a better fit
in the Fe line region. We also note the presence of a narrow feature
around 7.2--7.4~keV (6.5--6.7~keV in the observed frame). We will
discuss the possible origin of this feature in section
\ref{strangeline} below.

In addition to the interpretation that the inner disc is missing,
there are several possible scenarios that could explain the large
values obtained for $r_{\rm in}$ in our best-fitting models.  For
example, extreme ionization of the inner disc would lead to a nearly
featureless reflection spectrum that would be hard to disentangle from
the power law continuum. In fact, adding an inner reflector with $\xi$
fixed at $10^4\ \rm{erg\ cm\ s^{-1}}$ (the highest allowed value) to
our best-fitting reflection model results in an acceptable fit,
although the quality of the data is not good enough to constrain the
fit parameters. 

Another possible reason for the large value of $r_{\rm in}$ is that
the illumination of the disc results in an emissivity index that
deviates from our assumed value of $q=3$. The quality of the data does
not allow us to constrain this parameter if all other parameters are
left free. However, when fixing the other reflection parameters at
their time-averaged values, we find that the model with $r_{\rm in} =
1.24\ r_{\rm g}$ improves significantly with a flatter emissivity of
$q\sim 1$. Such a flat emissivity would be expected (at least in an
average sense) if the disc is illuminated from a great height ($\ga
30\ r_{\rm g}$) in a lamp post geometry (see also Vaughan et
al. 2004), or if the power-law emission originates in the base of a
jet, which is beamed away from the inner disc.

A contribution from a jet in 4C~+74.26 has previously been suggested
for the \asca observation of the source (Sambruna et al. 1999). In
order to model this scenario we add a second, hard power law to the
reflection models described above. The results of this are
inconclusive though, as we find that we cannot obtain any meaningful
constraints on the slope and normalization of the second power law.

\subsubsection{The narrow feature around 7.4~keV}
\label{strangeline}

As mentioned above, the residuals in Fig. \ref{refmodels} reveal a
narrow feature around 7.2--7.4~keV (6.5--6.7~keV in the observed
frame). Depending on the underlying model, this feature can be
modelled either as a 7.40~keV emission line or as a 7.17~keV
absorption line.

For the model consisting of a power law and a broad Gaussian, the best
fit is obtained with a narrow emission line at $7.40$~keV (the fit
improves by $\Delta \chi^2 =-8$ for 2 fewer degrees of freedom when
the line is added, and EW$=16\pm 11$~eV).  The same is true for the
Laor and reflection models in table \ref{results1} that have $r_{\rm
  in}$ as a free parameter. None of the best-fitting parameter values
change significantly when the line is added to these models.  An
absorption line at $7.17$~keV results in a very small improvement of
the fit for these models, and we especially note that the addition of
the line does not significantly affect the inclination. This is also
true if the absorption line is identified with Fe {\scriptsize XXVI}
at 6.97~keV and we add a Fe {\scriptsize XXV} line (rest energy
6.7~keV) with the same blueshift.

For the models where $r_{\rm in}=1.24\ r_{\rm g}$ and a narrow 6.4~keV line
is included, we find that the narrow feature can be equally well
fitted with a $7.17$~keV absorption line as with a $7.40$~keV emission
line. Specifically, in the case of the model which consists of a power
law, a Laor line and a 6.4~keV line, we find that the absorption line
improves the fit by $\Delta \chi^2 =-13$ for 2 fewer degrees of
freedom and has EW$=22^{+10}_{-9}$~eV, while the emission line results
in $\Delta \chi^2 =-14$ for 2 fewer degrees of freedom and has
EW$=16^{+10}_{-11}$~eV. No model that includes both an absorption and
an emission line can be found. Although the absorption line is an
equally good fit as the emission line in this case, we note that the
inclination goes up to about $55^{\circ}$ when the absorption line is
added. This is higher than the upper limit of $49^{\circ}$ placed by
Doppler boosting arguments (Pearson et al. 1992).

We have ruled out an instrumental origin for the lines, which suggests
that they (if real) are due to highly blueshifted Fe emission. For
example, identification of the 7.17~keV absorption line with Fe
{\scriptsize XXV} at 6.70~keV would imply an outflow velocity of about
$20\ 000\ \rm{km\ s^{-1}}$.  This extreme velocity together with the
relatively low significance of the lines suggests that they are
probably not real. However, it is interesting to note that Robinson et
al. (1999) detected a redshifted H$\alpha$ line in polarised light in
this source, indicating that an outflow with a velocity greater than
$5\ 000\ \rm{km\ s^{-1}}$ is present.

\subsection{The high-energy PIN spectrum}

4C +74.26 is clearly detected in the PIN up to about 30~keV and we
stress that this detection is robust against uncertainties in the
background model. When fitting the data we use the model for the
non-X-ray background provided by the HXD-team, and we account for the
contribution from the CXB as described in section \ref{observations}.
The cross normalization of the PIN with respect to XIS0 has been
reported to be about 1.14 for the HXD nominal position and
{\scriptsize V}2 data (Ishida, Suzuki \& Someya 2007). After adjusting
this value for the difference in normalization between XIS0 and our
added FI XIS spectrum, we adopt a PIN/FI XIS cross normalization of
1.12.

Fig. \ref{ratio} shows the PIN + FI XIS spectrum as a ratio to a power
law, fitted over the 2--4 and 7--10~keV energy ranges. In order to
illustrate the uncertainties in the PIN data we show the spectra
obtained by increasing and decreasing the background model by 3 per
cent (the one-sigma uncertainty of v2 background models). We note that
the study of the PIN light curves in section \ref{lightcurves} showed
that the background has probably been underestimated, suggesting that
the real value is closer to the lower (red) data points. This also
agrees with the predictions of the reflection models that were found
to fit the 2--10~keV data.  For the model where $r_{\rm in}$ is a free
parameter we get a best 2--30~keV fit of $\chi^2/\rm{d.o.f.} =
1561/1549$ if the background is increased by 2 per cent. Similarly,
for the model with $r_{\rm in}=1.24\ r_{\rm g}$, we get a best
chi-squared of $\chi^2/\rm{d.o.f.} = 1573/1548$ if the background is
increased by 1--2 per cent.

The relative contribution from the reflection component is usually
measured by the reflection fraction, which is the ratio of the total
reflected emission to the power-law emission. For an isotropic source
illuminating an infinite plane of gas the reflection fraction is 1. We
estimate the reflection fraction of our models by measuring the the
unabsorbed fluxes in the two components within {\scriptsize
  XSPEC}. For the two models presented above the reflection fractions
are 0.3 and 0.5, respectively, assuming that the power law emission
extends all the way up to the upper boundary for the {\scriptsize
  REFLION} model. On the other hand, assuming an upper cutoff for the
power law of 150~keV (as previously found for \emph {BepposSAX} data,
Grandi et al. 2006) we find slightly higher reflection fractions of
0.4 and 0.7.  In either case, it is clear that the reflection fraction
is less than 1, implying e.g. that the illumination is anisotropic,
that the disc is truncated or that an unreflected component (such as a
jet) contributes to the spectrum.

\subsection{Comparison with the Swfit and INTEGRAL data}

\begin{figure}
\begin{center}
  \rotatebox{270}{\resizebox{!}{80mm}{\includegraphics{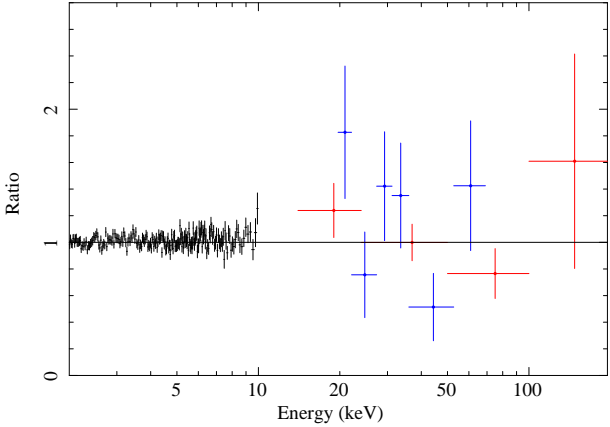}}}
  \caption{\small{The \suzaku FI XIS spectrum (in black), together
      with the {\it Swift} (red) and {\it INTEGRAL} (blue) spectra,
      shown as a ratio to the best-fitting 2--10~keV reflection
      model. The normalizations of the {\it Swift} and {\it INTEGRAL}
      data were allowed to vary freely with respect to the XIS
      spectrum.}}\label{ratio_swift_int}
\end{center}
\end{figure}

4C~+74.26 is included in both the {\it Swift} BAT 9-month AGN catalog
(Tueller et al. 2007) and the {\it INTEGRAL} general reference catalog
(Bodhagee et al. 2007). In order to further investigate the
high-energy properties of 4C~+74.26 we briefly investigate the spectra
from these catalogs in this section. The data from the BAT catalog are
publicly
available\footnote{http://swift.gsfc.nasa.gov/docs/swift/results/bs9mon/}
and the {\it INTEGRAL} spectrum was provided by the {\it INTEGRAL} AGN
team (the {\it INTEGRAL} data is also analysed in Molina et al. 2008,
in prep).

A simple power-law fit to the data yields $\Gamma=2.1^{+0.4}_{-0.3}$
for {\it Swift} (14--195~keV) and $\Gamma=2.2^{+0.7}_{-0.6}$ for {\it
  INTEGRAL} (20--100~keV). This should be compared with $\Gamma=1.8\pm
0.2$ in the 14--30~keV range for the \suzaku PIN detector. No clear
evidence for a cutoff of the power law is seen in any of the
spectra. Based on the power-law fits the inferred 14--30~keV fluxes
are $F_{14-30}=\rm{1.4\times 10^{-11}\ erg\ cm^{-2}\ s^{-1}}$ ({\it
  Swift}) $F_{14-30}=\rm{2.6\times 10^{-11}\ erg\ cm^{-2}\ s^{-1}}$
({\it INTEGRAL}) and $F_{14-30}=\rm{2.7\times 10^{-11}\ erg\ cm^{-2}\
  s^{-1}}$ (\suzaku PIN). As the observations are not simultaneous and
the source appears to be getting brighter (as shown in Table
\ref{longterm}), it is not surprising that the photon indices and
fluxes differ somewhat.

In order to check if the {\it Swift} and {\it INTEGRAL} data are
consistent with a reflection model, we fit them with our best-fitting
2--10~keV reflection model, leaving only the cross-normalizations
between the detectors as free parameter. As shown in
Fig. \ref{ratio_swift_int}, the shape of the high-energy spectra can
be well reproduced by the reflection model ($\chi^2/\rm{dof} =
1527/1524$ over the entire 2--195~keV range). The cross-normalizations
with respect to the FI XIS are found to be 0.6 for {\it Swift} and 0.9
for {\it INTEGRAL}. We also note that the high-energy data can be
equally well described with the reflection model with $r_{\rm
  in}=1.24\ r_{\rm g}$.

\subsection{Fits including the previous \xmmn data}
\begin{figure}
\begin{center}
\rotatebox{270}{\resizebox{!}{80mm}{\includegraphics{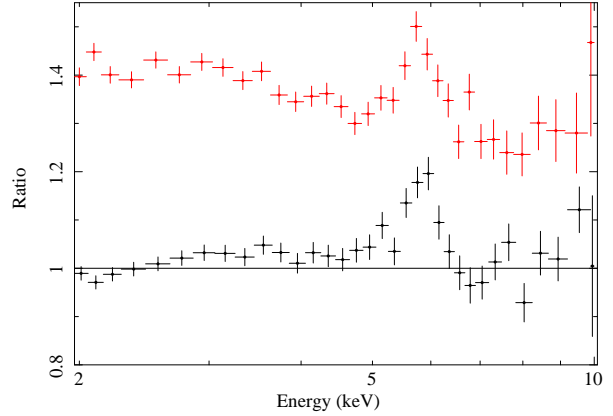}}}
\caption{\small{Comparison of the \xmmn EPIC pn (black) and
    \suzaku FI XIS (red) spectra. The plot shows the spectra as
    a ratio to a power law fitted to the \xmmn data, excluding
    the 4--7~keV energy band.}}\label{xmmcomp}
\end{center}
\end{figure}

\begin{table*}
\begin{center}
\begin{tabular}{lrrrrrrrrrr}
\hline\hline \\ [-4pt]
Model & \multicolumn{1}{c}{$\Gamma_{s}$} & \multicolumn{1}{c}{$\Gamma_{x}$}
& \multicolumn{1}{c}{$r_{\rm{in}}$} & \multicolumn{1}{c}{$i$}
& \multicolumn{1}{c}{Fe} & \multicolumn{1}{c}{$\xi$} 
&  \multicolumn{1}{c} {${\rm E_{nl}}$} 
& \multicolumn{1}{c} {${\rm EW}_{{\rm nl}, s}$} &  {${\rm EW}_{{\rm nl}, x}$}
& \multicolumn{1}{c}{$\chi^2$/dof} \\[4pt]  \hline\\[-6pt]

pl + blr*ref & $1.83^{+0.02}_{-0.02}$ & $1.72^{+0.01}_{-0.02}$ & $161^{+239}_{-117}$
& $31^{+12}_{-11}$ & $3.8^{+6.2}_{-2.2}$ & $30^{+21}$ &   &  &  & 2410/2538  \\

pl + blr*ref + nl  & $1.86^{+0.03}_{-0.04}$ & $1.76^{+0.05}_{-0.05}$
& $1.24^{\rm{f}}$ & $40^{+4}_{-5}$ & $2.5^{+7.5}_{-0.8}$ & $30^{+9}$ 
& $6.4^{+0.04}_{-0.02}$ & $28^{+7}_{-8}$   & $36^{+10}_{-10}$ & 2431/2537\\
\hline
\end{tabular}
\caption{\label{jointresults}\small{Combined 2--10~keV reflection fits
    to the \suzaku and \xmmn spectra. The model components are pl =
    power law, blr = {\scriptsize KDBLUR} (relativistic blurring), ref
    = {\scriptsize REFLION} (reflection model), nl = narrow Gaussian
    emission line ($\sigma$ fixed at 1~eV). $\Gamma$ is the photon
    index of the power law, $r_{\rm{in}}$ is the inner radius of the
    disc (in gravitational radii), $i$ is the inclination of the disc
    to the line of sight (in degrees), Fe is the iron abundance of the
    disc (relative to solar), $\xi$ is the ionization parameter of the
    disc (in ${\rm erg\ cm^{-2}\ s^{-1}}$), ${\rm E_{nl}}$ is the
    energy of the narrow line (in keV), ${\rm EW}_{{\rm nl}}$ is the
    equivalent width of the narrow line (in eV). Subscripts $s$ and
    $x$ denote values for the \suzaku and \xmmn spectra,
    respectively. All parameters were tied between the two spectra
    apart from the photon index and the normalization of the power
    law. Superscript f indicates that the parameter was frozen. The
    following parameters were fixed in all the fits: the outer radius
    of the disc (400 gravitational radii), the emissivity index of the
    disc (3), $\sigma$ of the narrow line (1~eV). All fits also
    included absorption fixed at the Galactic value.}}
\end{center}
\end{table*}

Fig. \ref{xmmcomp} shows a comparison of the \suzaku FI XIS spectrum
and the EPIC pn spectrum from the previous \xmmn observation of
4C~+74.26 (Ballantyne \& Fabian 2005). The spectra are shown as a
ratio to a power-law fitted to the \xmmn data, excluding the 4--7~keV
energy range. It is clear from the plot that the continuum is somewhat
steeper in the \suzaku observation ($\Delta \Gamma \sim 0.1$), while
the Fe line profiles are very similar in both spectra.

In order to to try and find a spectral model for both observations we
performed simultaneous fits to the spectra, using the two different
reflection fits obtained for the \suzaku data as a starting point.
Experimentation showed that very good joint fits could be obtained for
both models if all parameters apart from the slope and the
normalization of the power law were tied between the two spectra.  The
results of the fits are presented in Table \ref{jointresults}. We see
that the model where $r_{\rm{in}}$ is a free parameter is still a
somewhat better fit than the model with $r_{\rm{in}}=1.24\ r_{\rm g}$
($\chi^2/\rm{d.o.f.}=2410/2538$ compared to 2431/2537). In both models
the photon index of the power law is 0.1 steeper for \suzaku than for
\xmmn. The only tied parameters that differ significantly from the
values obtained when fitting only the \suzaku data are $r_{\rm{in}}$,
$i$ and Fe of the first model. The values of these parameters have all
increased (especially note that the inclination now has a more
reasonable value of $31^{+12}_{-11}\deg$) but they are all still
consistent (within the errors) with the \suzaku fits. As in the case
of the \suzaku fits, we also find that both models agree well with the
high-energy PIN data, and that the best agreement is found if we
assume that the background model has been underestimated by 2--3 per
cent.

We stress that the reflection normalization is tied in these fits and
that letting it vary independently for the two spectra does not
improve the quality of the fits. The same is true for the
normalization of the narrow 6.4~keV line.  We thus conclude that the
difference between the \suzaku and \xmmn observations can be well
described in terms of a power law that has increased in normalization
and also become slightly steeper. 

As a final comment we note that there is evidence for a narrow 6.2~keV
line in the \xmmn spectrum (EW~$\sim$20~eV, as previously noted by
Ballantyne \& Fabian 2005) but not in the \suzaku spectrum. The origin
of this line is unclear but Ballantyne \& Fabian (2005) suggested that
it and the 6.4~keV line could form the red and blue horns of a
diskline originating in the outer disc.

\section{Discussion}
\label{discussion}

\subsection{Spectral variability}

Due to the long duration of the \suzaku observation, we have for the
first time detected variability in the light curve of 4C~+74.26. We
find that the source brightens by about 20 per cent during the
observation, implying a doubling time of $\sim 1$~Ms. There is
evidence for a positive correlation between the hardness ratio and the
count rate, although this is only significant at the 99 per cent level
in the case of XIS1.  We also find that the rms variability increases
with energy up to about 4~keV, at which point it flattens out and
shows evidence of decreasing. This spectral variability is rather
different from that observed in typical radio-quiet Seyfert galaxies,
which generally exhibit spectral softening with increasing count rate
and rms variability decreasing with energy above $\sim 1$~keV
(e.g. Papadakis et al. 2002; Markowitz \& Edelson 2004).

One possible explanation for the spectral variability observed in
4C~+74.26 is that a jet contributes to the spectrum. This is possible
even if the inclination of the jet to the line of sight is fairly
large, as the base of the jet may not be well collimated. In this
picture, our measured photon index of $\sim 1.8$ would be due the
combined contributions from a hard component associated with the jet,
and a softer component associated with Comptonized emission from the
corona, as previously suggested for an \asca observation of the source
(Sambruna et al. 1999). The fact that the source is harder when it is
brighter would then indicate that the relative contribution from the
jet is higher when the flux increases, assuming that the intrinsic
slopes of the two power laws do not vary significantly. Such a model
could in principle be tested by investigating the difference spectrum
between high- and low-flux spectra. However, due to the limited
amplitude of the variability in 4C~+74.26 we find that the difference
spectrum is not of sufficient quality to discriminate between models.

\subsection{Spectral modelling}

In our spectral analysis of the \suzaku data of 4C~+74.26 we clearly
detect a broad Fe line, as previously seen by \xmmn (Ballantyne \&
Fabian 2005).  Our best fits to the Fe line profile suggest that the
line originates outside about 50~$r_{\rm g}$ and that the inclination to the
line of sight is around $20^{\circ}$. This inclination implies that
the source is around 3~Mpc across, which is extremely large even for a
giant radio galaxy.  Although it is possible that the source actually
is this large, we note that an almost equally good fit can be obtained
if the inclination is fixed at a larger value. For this scenario the
inner radius of emission is outside 100~$r_{\rm g}$. We also managed to find
a good fit with a model in which $r_{\rm{in}}$ was fixed at $1.24\
r_{\rm g}$. This model has the advantage of a higher inclination of around
$40^{\circ}$, but is a worse fit in the Fe line region and also relies
on the presence of a narrow Fe~K$\alpha$ line, which is not required
in any other model. Overall it seems like the first model is a better
approximation of the real conditions.

The perhaps simplest explanation for these findings is that the inner
accretion disc is missing, as suggested for many other radio-loud AGN
(e.g. Eracleous et al. 2000). It is also possible, however, that the
inner disc is highly ionized or not strongly illuminated. The latter
explanation seems more likely, as the former model would require a
very sharp transition from an extremely ionized to a nearly neutral
disc. A plausible explanation for why the inner disc is not strongly
illuminated is that the corona, where the power law originates, is
located high above the disc.

The coronae of AGN are widely believed to be powered by magnetic
fields in their accretion discs. As magnetic fields are thought to be
important for jet formation (e.g. Blandford \& Payne 1982), it is
likely that the properties and structure of the magnetic fields in
radio-quiet and radio-loud objects are different. The strong
reflection features often seen in radio-quiet AGN indicate that their
inner discs are strongly illuminated, implying that the magnetic
energy is dissipated close to the disc. It is conceivable that the
magnetic field structure in radio-loud objects leads to the energy
being dissipated at larger heights, possibly in a region moving away
from the disc, thus resulting in less illumination of the inner disc
and weaker reflection features.

It is also possible that emission from the jet itself contributes to
the spectrum, as tentatively suggested by the variability properties
described above. We explicitly tested for the presence of a jet in the
\suzaku spectrum by adding a second, hard power law to the power law +
reflection model. The results were inconclusive, however, as we found
that we could not constrain the properties of the second power law. A
much longer observation would be needed to put constraints on the
possible jet component in 4C~+74.26.

The difference between the 2007 \suzaku and 2004 \xmmn spectra can be
well modelled with a power law that has increased in normalization and
become slightly steeper. The reflected emission is consistent with
staying constant between the observations, and the best-fitting disc
parameters in the joint fits are similar to those found when only
fitting the \suzaku data. A possible explanation for this behaviour is
that the increase in flux between the two observation is mainly due to
the (unreflected) emission from the jet increasing, in combination
with the slope of the coronal and/or jet power laws steepening
somewhat.

\section{Summary and conclusions}
\label{conclusions}

We have analysed a 92~ks exposure \suzaku observation of the radio
quasar 4C~+74.26. Our main results can be summarized as follows:

\begin{itemize}

\item{The 2--10~keV flux measured by \suzaku is $\rm{{3.24\times
        10^{-11}\ erg\ cm^{-2}\ s^{-1}}}$, which is the highest flux
    observed to date.  The source has gradually brightened by about a
    factor of two since it was first observed by \asca in 1996}.\\

\item{Due to the long observation, we have for the first time detected
    variability in the light curve of 4C~+74.26. We find that the
    count rate increases by about 20 per cent during the
    observation.\\}

\item{There is evidence for spectral hardening as the count rate of
    the source increases.  We also find that rms variability increases
    with energy up to about 4~keV, where it flattens out and shows
    some evidence of decreasing.\\}

\item{We clearly detect a broadened Fe~K$\alpha$ line. When fitted
    with a simple Gaussian, it has $EW=85$~eV and $\sigma = 0.24$~keV.\\}

\item{The entire 2-30~keV energy range can be well fitted with a power
    law ($\Gamma \approx 1.8$) and near-neutral reflection from an
    accretion disc. No reflected emission inside about 50~$r_{\rm g}$ is
    required by the Fe line profile, although a model with
    $r_{\rm{in}}= 1.24\ r_{\rm g}$ provides a good fit if a narrow
    Fe~K$\alpha$ line is included. The reflection fraction is found to
    be in the range $\sim 0.3$--0.7, depending on the model.\\}

\item{The high-energy \emph{Swift} BAT and \emph{INTEGRAL} spectra
    show a somewhat steeper continuum slope ($=\Gamma \sim 2.1-2.2$)
    but can nevertheless be well described with the power-law +
    reflection models described above.\\}

\item{The difference between the \suzaku and \xmmn 2--10~keV spectra
    can be well explained with a power law that has increased in
    normalization and become slightly steeper.}

\end{itemize}

The spectral variability properties described above can be explained
in a model where the relative contribution from a hard jet increases
with increasing flux. However, because of the limited variability
amplitude and spectral quality, we are not able to constrain the
possible contribution from a jet in 4C~+74.26. The best-fitting
reflection model implies that the inner disc is either truncated at
large radii or not strongly illuminated. We suggest that the latter
scenario may be due to the power-law source being located high above
the disc, e.g. as a result of the magnetic field structure.

\section{Acknowledgements}
We thank Alessandra De Rosa for providing the {\it INTEGRAL} spectrum.
JL thanks Corpus Christi College, the Isaac Newton Trust and STFC. ACF
thanks the Royal Society for support. D. R. B. is supported by the
University of Arizona Theoretical Astrophysics Program Prize
Postdoctoral Fellowship.

{}

\end{document}